# TWO-DIMENSIONAL PHONONIC CRYSTALS WITH ACOUSTIC-BAND NEGATIVE REFRACTION

Hossein Sadeghi[1] and Sia Nemat-Nasser[2]
[1]Karagozian & Case, 700 N. Brand Blvd., Suite 700, Glendale, CA 91203
[2]Department of Mechanical and Aerospace Engineering, UC San Diego, La Jolla, CA 92093

**ABSTRACT**

A two-dimensional phononic crystal (PC) can exhibit longitudinal-mode negative energy refraction on its lowest (acoustical) frequency pass band. The effective elastodynamic properties of a typical PC are calculated and it is observed that the components of the effective density tensor can achieve negative values at certain low frequencies on the acoustical branches for the longitudinal-mode pass-band, and that negative refraction may be accompanied by either positive or negative effective density. Furthermore, such a PC has a high anisotropy ratio at certain low frequencies, offering potential for application to acoustic cloaking where effective material anisotropy is essential.

**INTRODUCTION**

Phononic crystals (PCs) are artificial materials with specially designed microstructure to control stress waves [1]. Microstructure of a PC can be designed to achieve negative energy refraction at the interface of the PC and a homogenous medium at certain frequency ranges. This feature can be used to focus stress waves in a focal point in order to make flat acoustic lens for applications like ultrasound imaging. Furthermore, due to recent advances in transformational acoustics, which makes acoustic cloaking achievable [2], dynamic homogenization has become a powerful tool for microstructural design of the cloak [3]. This demands further understanding of dynamic homogenization techniques and their limitations. In this paper, a mixed variational method [4] is used together with dynamic homogenization [5] to study energy refraction in two-dimensional PCs. Equifrequency surfaces (EFS) of a two-dimensional PC made of epoxy matrix with steel inclusions are calculated using mixed variational method. Vectors of group velocity are studied and energy refraction at the interface of a homogenous half-space and the PC is investigated. Frequency-dependent effective elastodynamic properties of the PC are obtained through micromechanical method to study overall behavior of the PC at different frequencies.

In the recent years, there have been many efforts to study negative energy refraction behavior in PCs. Yang et al. [6] presented a combined experimental and theoretical study of negative refraction in three-dimensional PCs. They showed that three-dimensional PCs can be used to focus a diverging ultrasound beam into a narrow focal spot. Li et al. [7] used the multiple scattering technique and studied negative energy refraction of acoustic waves in two-dimensional phononic crystals. They showed that local resonance mechanism brings on a group of flat bands in low frequency region which provides two EFS's close to circular leading to negative refraction. Croenne et al. [8] presented experimental evidence of negative refraction of longitudinal waves in two-dimensional PCs with a solid matrix. They fabricated a PC made of triangular arrangements of steel rods embedded in epoxy and carried out an experiment on a prism-shaped PC inside an epoxy block and observed negative refraction experimentally. Nemat-Nasser [9, 10] studied anti-plane shear wave propagation in one- and two-dimensional PCs using a mixed variational formulation. He showed that negative energy refraction can be accompanied by positive phase-velocity refraction, and positive energy refraction can be accompanied by negative phase-velocity refraction.

## MIXED VARIATIONAL METHOD

Consider a two-dimensional PC with rectangular inclusions and a unit cell shown in Figure 1. For a unit cell of the PC with the edges defined by $h^\gamma$ ($\gamma = 1, 2$), the periodicity condition can be expressed as

$$C_{jkmn}(x) = C_{jkmn}(x + m'h^\gamma)$$
$$\rho(x) = \rho(x + m'h^\gamma) \quad (1)$$

where $x$ is the position vector, $\rho(x)$ is the density, $C_{jkmn}(x)$ is the elasticity tensor, and $m'$ can be any integer.

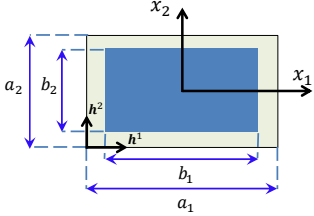

Figure 1: Unit cell of a two-dimensional PC with rectangular inclusions

Consider harmonic waves with frequency $\omega$ and write the governing equation and stress-displacement relation as

$$\sigma_{jk,k} + \rho\omega^2 u_j = 0$$
$$\sigma_{jk} = C_{jkmn} u_{m,n} \quad (2)$$

where $u_j$ and $\sigma_{jk}$ are components of displacement and stress tensors, respectively. For harmonic waves with wave vector $q$ the Bloch boundary conditions have the following form

$$u_j(x + h^\gamma) = u_j(x) e^{iq \cdot h^\gamma}$$
$$t_j(x + h^\gamma) = -t_j(x) e^{iq \cdot h^\gamma} \quad (3)$$

where $t$ is the traction vector. It can be shown that the eigenvalues of the problem can be found by rendering the following functional stationary

$$\lambda_N = \frac{\langle \sigma_{jk}, u_{j,k}\rangle + \langle u_{j,k}, \sigma_{jk}\rangle - \langle D_{jkmn}\sigma_{jk}, \sigma_{mn}\rangle}{\langle \rho u_j, u_j\rangle} \quad (4)$$

In mixed variational method, components of stress and displacement tensors are to varied independently and are replaced by their Fourier series expansions. The problem is reduced to an eigenvalue problem which is then solved to find the dispersion relation of the PC.

## MICROMECHANICS FORMULATION

A homogenization technique is used for calculation of the overall elastodynamic properties of two-dimensional PCs based on micromechanical method [13]. The solution to the equations of motion for a two-dimensional elastic composite can be expressed as the sum of the volume average and a disturbation term due to heterogeneities in the unit cell as

$$\hat{\phi} = \phi^0 + \phi^d \quad (5)$$

where $\hat{\phi}$ represents any of the field variables, stress ($\hat{\sigma}$), strain ($\hat{\varepsilon}$), momentum ($\hat{p}$) or velocity ($\hat{u}$). For Bloch type waves the field variables can be written as

$$\hat{\phi}(x, t) = Re[\phi(x) \exp[i(q \cdot x - \omega t)]] \quad (6)$$

where $\phi$ represents the periodic parts of the field variables. The heterogeneous unit cell is replaced by a homogenous one with uniform density $\rho^0$ and compliance $\mathbf{D}^0$. Eigenstrains, $\mathbf{E}(x)$, and eigenmomentums, $\mathbf{P}(x)$, are introduced such that the pointwise values of the field variables are the same as the original heterogeneous solid. Therefore, the consistency conditions can be expressed as

$$\boldsymbol{\varepsilon} = \mathbf{D}:\boldsymbol{\sigma} = \mathbf{D}^0:\boldsymbol{\sigma} - \mathbf{E}$$
$$\boldsymbol{p} = \rho \dot{\boldsymbol{u}} = \rho^0 \dot{\boldsymbol{u}} - \mathbf{P} \quad (7)$$

Equations (7) are substituted in the governing equations (2) and the field variables are replaced with their Fourier series expansion. The consistency conditions and governing equations are then averaged over the unit cell and the integrals are replaced with their equivalent finite sums (see [5] for more details). Doing some simplifications the overall constitutive equations of a 2-D PC is obtained which has the following form

$$\langle \boldsymbol{\varepsilon}\rangle = \overline{\mathbf{D}}:\langle \boldsymbol{\sigma}\rangle + \overline{\mathbf{S}}^1 \cdot \langle \dot{\boldsymbol{u}}\rangle$$
$$\langle \boldsymbol{p}\rangle = \overline{\mathbf{S}}^2:\langle \boldsymbol{\sigma}\rangle + \overline{\boldsymbol{\rho}} \cdot \langle \dot{\boldsymbol{u}}\rangle \quad (8)$$

where $\langle\boldsymbol{\varepsilon}\rangle$, $\langle\boldsymbol{\sigma}\rangle$, $\langle\dot{\boldsymbol{u}}\rangle$, $\langle\boldsymbol{p}\rangle$, $\overline{\mathbf{D}}$, and $\overline{\boldsymbol{\rho}}$ are the average strain, average stress, average velocity, average momentum, effective compliance tensor, and effective density tensor, respectively; while $\overline{\mathbf{S}}^1$ and $\overline{\mathbf{S}}^2$ are the coupling terms. It can be seen that in this formulation, mean stress is coupled not only to mean strain but also to mean velocity, and mean momentum is likewise coupled not only to mean velocity but also to mean strain. It should be noted that, the effective compliance and effective density tensors are in general anisotropic.

## RESULTS AND DISCUSSION

Consider a two-dimensional PC with square unit cell with dimensions given by $a_1 = a_2 = 3 \, cm$, and square inclusions with dimensions given by $b_1 = b_2 = 1 \, cm$. The matrix is made of epoxy and the inclusions are made of steel. Figure 2 shows the band structure of the PC for the first five modes. Figure 3 shows the comparison between the band structures obtained from mixed variational method and micromechanical method. It can be seen that as the frequency increases, the difference between the results obtained from these two methods increases.

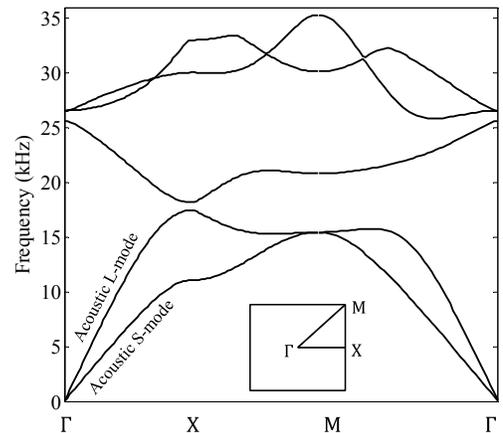

Figure 2: Frequency band structure of the two-dimensional PC

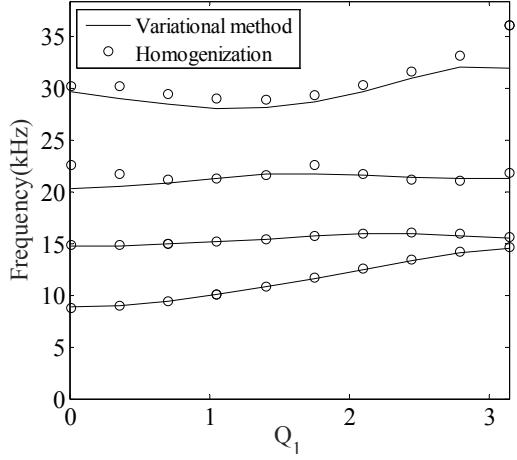

Figure 3: Comparison between the band structures calculated from mixed variational method and micromechanical method

Figure 4 shows EFS of the composite together with vectors of group velocity for the acoustic longitudinal band and it is observed that depending on the value and direction of the wave vector, components of the group velocity and phase velocity vectors can be in the same direction (positive refraction) or opposite directions (negative refraction). For example, at $\boldsymbol{Q}_A$=(2.307, 2.307) the EFS contour is quasi-circular with antiparallel phase and group velocities resulting in negative energy refraction.

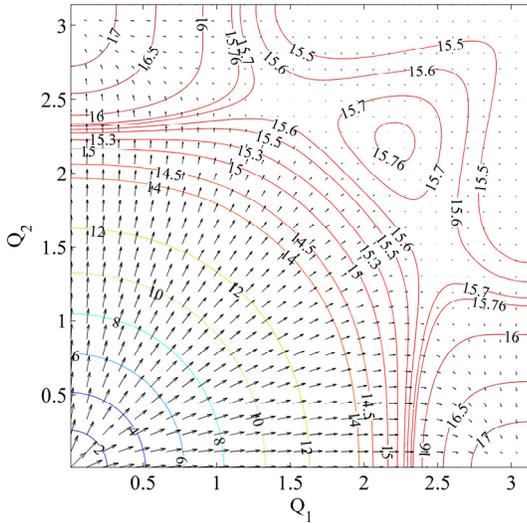

Figure 4: Contours of EFS for the 2-D PC over the acoustic L-mode

Furthermore, the contours of components of the effective compliance and effective density are calculated and it is observed that negative energy refraction can be accompanied by either positive or negative effective properties (see Figure 5). Figure 6 shows the values of the effective density, effective compliance, and effective coupling term along the $Q_1$-axis for a fixed value of $Q_2$ over the acoustic L-mode. Negative energy refraction in this PC can be used to focus acoustic/pressure waves in a focal point in order to make flat acoustic lens for applications like ultrasound imaging, or to focus high intensity ultrasound acoustic/pressure waves for cancer treatment. In addition, it is observed that even though the unit cell and inclusions are square and symmetric, the homogenized medium is anisotropic. More importantly, the anisotropy ratio at some frequencies is very high which could be used for design of the cloak for acoustic cloaking where unit cells with high anisotropy are essential.

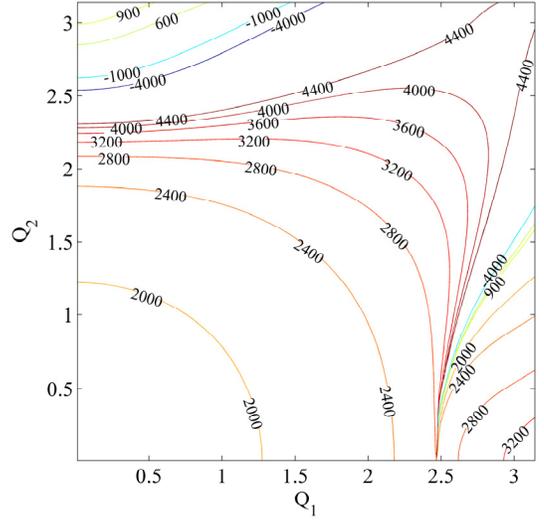

Figure 5: Contours of effective density for the 2-D PC over the acoustic L-mode

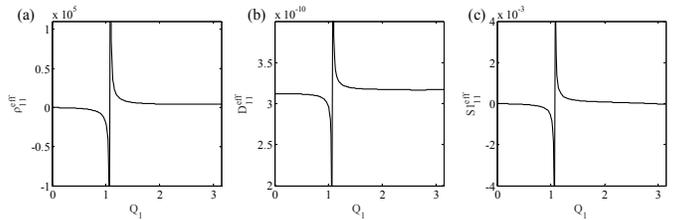

Figure 6: Values of (a) $\bar{\rho}_{11}$, (b) $\bar{D}_{11}$, and (c) $\bar{S}_{11}^{1}$ along the $Q_1$-axis for a fixed value of $Q_2$ over the acoustic L-mode

## CONCLUSION

Stress wave propagation in a two-dimensional PC made of epoxy matrix with steel inclusions is studied. It is observed that over the acoustic longitudinal mode components of the effective density tensor can become negative. In addition, vectors of group velocity are studied and it is observed that negative energy refraction can occur at some frequencies over the acoustic longitudinal mode, which can be accompanied by either positive or negative effective density. Furthermore, it is observed that the PC exhibits a high anisotropy ratio for the effective density at some frequencies over the acoustic longitudinal mode, which could be used for design of cloak for acoustic cloaking where unit cells with high anisotropy are essential.

## ACKNOWLEDGMENTS

This research has been conducted at the Center of Excellence for Advanced Materials (CEAM) at the University of California, San Diego, under DARPA Grant RDECOM W91CRB-10-1-0006 to the University of California, San Diego.

## REFERENCES

[1] S. Sia Nemat-Nasser, H. Sadeghi, A. V. Amirkhizi and A. Srivastava, "Phononic Layered Composites for Stress-wave Attenuation," *Mechanics Research Communications,* vol. 68, p. 65–69, 2015.

[2] G. W. Milton, M. Briane and J. R. Willis, "On cloaking for elasticity and physical equations with a transformation invariant form," *New Journal of Physics,* vol. 8, no. 10, p. 248, 2006.

[3] D. Torrent and J. Sanchez-Dehesa, "Acoustic cloaking in two dimensions: a feasible approach," *New Journal of Physics,* vol. 10, no. 6, p. 063015, 2008.

[4] S. Nemat-Nasser and S. Minagawa, "Harmonic waves in layered composites: comparison among several schemes," *Journal of Applied Mechanics,* vol. 42, p. 699, 1975.

[5] A. Srivastava and S. Nemat-Nasser, "Overall dynamic properties of three-dimensional periodic elastic composites," *Proceedings of the Royal Society A: Mathematical, Physical and Engineering Science,* vol. 468, no. 2137, pp. 269-287, 2012.

[6] S. Yang, J. Page, Z. Liu, M. Cowan, C. Chan and P. Sheng, "Focusing of sound in a 3D phononic crystal," *Physical review letters,* vol. 93, no. 2, p. 24301, 2004.

[7] J. Li, Z. Liu and C. Qiu, "Negative refraction imaging of solid acoustic waves by two-dimensional three-component phononic crystal," *Physics Letters A,* vol. 372, no. 21, pp. 3861-3867, 2008.

[8] C. Croenne, E. Manga, B. Morvan, A. Tinel, B. Dubus, J. Vasseur and A. Hladky-Hennion, "Negative refraction of longitudinal waves in a two-dimensional solid-solid phononic crystal," *Physical Review B,* vol. 83, no. 5, p. 054301, 2011.

[9] S. Nemat-Nasser, "Anti-plane shear waves in periodic elastic composites: band structure and anomalous wave refraction," *Proc. R. Soc. A,* vol. 471, no. 20150152, 2015.

[10] S. Nemat-Nasser, "Refraction characteristics of phononic crystals," *Acta Mechanica Sinica,* vol. 31, no. 4, p. 481–493, 2015.

[11] S. Nemat-Nasser, F. Fu and S. Minagawa, "Harmonic waves in one-, two-and three-dimensional composites: Bounds for eigenfrequencies," *International Journal of Solids and Structures,* vol. 11, no. 5, pp. 617-642, 1975.

[12] J. Willis, "Exact effective relations for dynamics of a laminated body," *Mechanics of Materials,* vol. 41, no. 4, pp. 385-393, 2009.

[13] H. Sadeghi, "Microstructurally Controlled Composites with Optimal Elastodynamic Properties," University of California San Diego, San Diego, 2016.